\documentclass[%
 reprint,
 amsmath,amssymb,
 aps,
]{revtex4-1}

\usepackage{graphicx}
\usepackage{dcolumn}
\usepackage{bm}

\newcommand{\tilt}{\theta_{\text{tilt}}}
\newcommand{\hys}{\Delta \theta}
\newcommand{\tapp}{\theta_\text{app}}
\newcommand{\tadv}{\theta_\text{adv}}
\newcommand{\trec}{\theta_\text{rec}}
\newcommand{\tadvrec}{\theta_\text{adv, rec}}
\newcommand{\dcos}{\Delta \cos \theta}
\newcommand{\Ko}{K_{\text{o}}}

\DeclareMathOperator\arcsinh{arcsinh}

\begin{document}


\title{Origins of liquid-repellency on structured, flat, and lubricated surfaces}

\author{Dan Daniel}
 \email{daniel@imre.a-star.edu.sg}
\author{Jaakko V. I. Timonen}
\author{Ruoping Li}
\author{Seneca J. Velling}
\author{Michael J. Kreder}
\author{Adam Tetreault}
\author{Joanna Aizenberg}
 \email{jaiz@seas.harvard.edu}
\affiliation{John A. Paulson School of Engineering and Applied Sciences, Harvard University, Cambridge, MA 02138, USA}

    
\begin{abstract}
There are currently three main classes of high-performance liquid-repellent surfaces: micro-/nano-structured lotus-effect superhydrophobic surfaces, flat surfaces grafted with `liquid-like' polymer brushes, and various lubricated surfaces. Despite recent progress, the mechanistic understanding of the differences in droplet behavior on such surfaces is still under debate. We measured the dissipative force acting on a droplet moving on representatives of these classes at different velocities $U$ = 0.01--1 mm/s using a cantilever force sensor with sub-$\mu$N accuracy, and correlated it to the contact line dynamics observed using optical interferometry at high spatial (micron) and temporal ($<$ 0.1s) resolutions. We find that the dissipative force---due to very different physical mechanisms at the contact line---is independent of velocity on superhydrophobic surfaces, but depends non-linearly on velocity for flat and lubricated surfaces. The techniques and insights presented here will inform future work on liquid-repellent surfaces and enable their rational design. 
\end{abstract}

\maketitle

\begin{figure}
\includegraphics[]{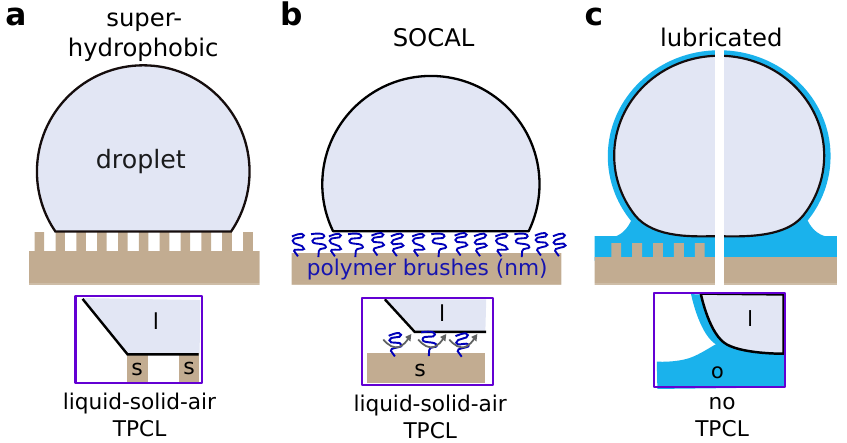} 
\caption{\label{fig:classes} Schematics of liquid-repellent surfaces. a, Structured lotus-effect superhydrophobic (SH) surfaces. b, Flat surfaces grafted with polymer brushes, dubbed Slippery Omniphobic Covalently Attached Liquid (SOCAL) surfaces. c, Structured (left) or flat (right) lubricated surfaces. The droplet is shown with a lubricant cloaking layer, which is usually the case for low-surface-tension lubricants (see Supplementary Fig. S1).}
\end{figure}

In Nature, the ability to repel water is often a matter of life and death. For example, insects must avoid getting trapped by falling raindrops and plants need to keep their leaves dry for efficient gas exchange through the stomata \cite{Smith1989, darmanin2015superhydrophobic}. Similarly, the tendency of water and complex fluids, such as blood and oil, to stick to surfaces pose many problems in industries, ranging from fuel transport, biomedical devices to hydrodynamic drag in ships \cite{quere2008wetting, bocquet2011smooth}. Hence, there is a huge interest in developing liquid-repellent materials. To achieve this, there are three main approaches. Firstly, hydrophobic micro-/nano-structures can be designed on the surface to maintain a stable air layer, minimizing contact between the applied liquid and the solid, i.e. lotus-effect superhydrophobic (SH) surfaces (Fig.~\ref{fig:classes}a) \cite{quere2008wetting, reyssat2010dynamical}. Secondly, a flat surface can be grafted with nanometer-thick `liquid-like' polymer brushes (Fig.~\ref{fig:classes}b); the resulting surface, known as Slippery Omniphobic Covalently Attached Liquid (SOCAL), is able to repel a wide variety of liquids, including low-surface-tension alkanes \cite{krumpfer2010contact, cheng2012statically, wang2016covalently}. Finally, a suitable lubricant oil can be added to the surface, which can be structured as is the case for Slippery Liquid-Infused Porous Surfaces (SLIPS) \cite{lafuma2011slippery, wong2011bioinspired} or flat as is the case of lubricant-infused organogels \cite{urata2015self, cui2015dynamic} (Fig.~\ref{fig:classes}c); any liquid can then easily be removed, as long as there is a stable intercalated lubricant layer \cite{smith2013droplet, Dan2016oleoplaning, keiser2017drop, smith2013droplet, schellenberger2015direct}. 

While the behaviors of liquids on each of these three surface types have been studied separately, there have been few attempts to compare their relative performance or explain their unique characteristics. In this letter, we will elucidate the origin of liquid repellencies for the three surface classes and how the details of the liquid-solid-air three phase contact line (TPCL) at the droplet's base---or the absence of TPCL in the case of lubricated surfaces---cause qualitatively different behaviors.

\begin{table}[b]
\caption{\label{tab:angle} Typical contact angle values for a water droplet}
\begin{ruledtabular}
\begin{tabular}{cccc}
	Surface &$\tapp$ &$\hys$ &$\dcos$\\
	\hline
	SOCAL &$\text{90--110}^{\circ}$ &$\text{1--10}^{\circ}$ &0.02--0.2 \\
	Lotus-effect &$>150^{\circ}$ &$\text{2--10}^{\circ}$ &0.02--0.1 \\
	Lubricated &$\text{90--110}^{\circ}$ &$\text{1--5}^{\circ}$ &0.02--0.05
\end{tabular}
\end{ruledtabular}
\end{table} 

To quantify and compare the liquid repellencies of these surfaces, previous work generally reports the static apparent contact angle $\tapp$ and the contact angle hysteresis $\hys = \tadv - \trec$, where $\tadvrec$ are the advancing and receding contact angles measured optically from the side (See Table \ref{tab:angle} and Supplementary Tables S1 and S2 for typical values) \cite{de2013capillarity}. The dissipative force acting on the moving droplet $F_{d}$ is related to the contact angle hysteresis $\dcos = \cos \trec - \cos \tadv$ by the Furmidge's  relation:    
\begin{equation} \label{eq:hys}
\begin{split}
F_{d} &= 2a\gamma \dcos
\text{,}
\end{split}
\end{equation}
where $a$ and $\gamma$ are the base radius and the surface tension, respectively \cite{de2013capillarity, furmidge1962}. For the three surface classes, a water droplet is reported to have similar hysteresis value of $\hys < 10^{\circ}$; however, for these studies, the exact experimental conditions---in particular, the speed of the contact line---are often not controlled, which is important because $F_{d}$ (and therefore $\hys$ and $\dcos$) can depend on the droplet's speed $U$ \cite{snoeijer2013moving, blake2006physics}. Moreover, there are technical challenges to contact angle measurements: $\theta$ is difficult to determine accurately when its value is too high $> 170^{\circ}$ (SH surfaces) or when obscured by a wetting ridge (lubricated surfaces) \cite{srinivasan2011assessing,  korhonen2013reliable, schellenberger2015direct}.  

\begin{figure}
\includegraphics[]{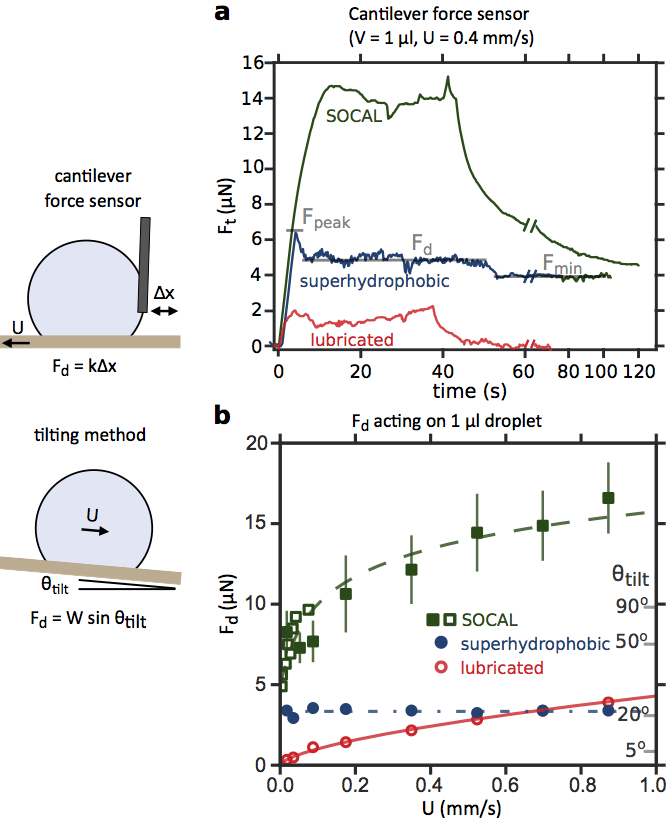} 
\caption{\label{fig:force} a, Characteristic force curves for a water droplet moving on the three surface classes measured using a cantilever force sensor. The motor (to move the substrate) was started at time $t$ = 0 s and stopped at $t \approx$ 50 s. b, $F_{d}$ for 1 $\mu$l water droplets moving at speeds $U$ = 0.01--1 mm/s on SH, SOCAL, and lubricated surfaces (filled circles, filled squares, and empty circles, respectively). $U$ of droplets tilted at different $\tilt$ = 25--90$^{\circ}$ on the same SOCAL surface and hence subjected to different $F_{d} = W \sin \tilt$ are shown on the same plot (empty squares). Each data point has three repeats with $\Delta F_{d} <$ 0.2 $\mu$N, unless otherwise indicated by error bars.}
\end{figure}

To better characterize the differences between the three surfaces, we measured $F_{d}$ directly using a cantilever force sensor, as described in our previous work (Fig.~\ref{fig:force}) \cite{Dan2016oleoplaning, pilat2012dynamic, gao2017drops}. The force acting $F_{t}(t)$ acting on the droplet moving at a controlled speed $U$ was inferred from the deflection $\Delta x(t)$ of the capillary tube attached to the droplet: $F = k \Delta x$, where $k = \text{5--25}$ mN/m for tube-lengths $L$ = 6--9 cm. 

\begin{figure*}
\includegraphics[]{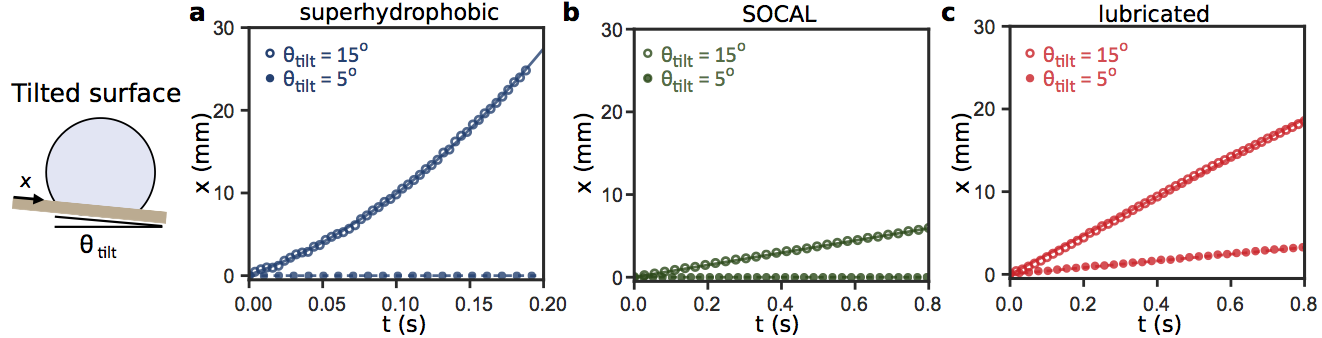} 
\caption{\label{fig:droplet_motion} Droplet motion on tilted substrates with $\tilt = 5^{\circ}$ and $15^{\circ}$ for the different surface classes. Depending on whether a droplet is moving with a constant speed $U$ or constant acceleration $a$, the displacement $x$ varies linearly or quadratically with $t$, respectively.}
\end{figure*}

Fig.~\ref{fig:force}a shows the characteristic force curves for the three surface classes. $F_{d}$ was taken to be the long-time average of $F_{t}$ once it had reached a steady state. For a lotus-effect surface decorated with micropillars (hexagonal array, diameter $d$ = 16 $\mu$m, pitch $p$ = 50 $\mu$m, and height $h_{\text{p}}$ = 30 $\mu$m), the force required to jumpstart the motion $F_{\text{peak}} = $ 6.6 $\mu$N is larger than the force to maintain the motion $F_{d} = 5.0 \pm 0.2$ $\mu$N, reminiscent of the static and kinetic friction forces between two solid surfaces \cite{gao2017drops}. In contrast, for lubricated and SOCAL surfaces (see Supplementary section S2 for details on sample preparation), there is no distinct $F_{\text{peak}}$. At time $t \approx 50$ s, the droplet motion was stopped, and for a lubricated surface, the cantilever returned to its original position ($\Delta x = 0$) and $F_{t} \rightarrow 0$; in contrast, for SOCAL and lotus-effect surfaces, $F_{t}$ did not return to zero, but reached a finite value $F_{\text{min}}$ ($F_{t} \rightarrow F_{\text{min}} > 0$), albeit with different $F_{\text{min}}$ and decay times. 

As $U$ was varied in the range of 0.01--1 mm/s, we found that $F_{d}(U)$ acting on a 1 $\mu$l water droplet exhibits different functional forms for the different surface types, suggesting different mechanistic origins for the liquid-repellency (Fig.~\ref{fig:force}b). Firstly, there is a minimum force required to move the water droplet on SH and SOCAL surfaces, $F_{\text{min}}$ = 4 and 5 $\mu$N, respectively; in contrast, $F_{d} \rightarrow 0$  as $U \rightarrow 0$ for similar water droplet on an ideal lubricated surface, i.e. $F_{\text{min}}$ = 0. Secondly, $F_{d}$ is independent of $U$ for SH surfaces (dash-dot line, Fig.~\ref{fig:force}b), but has a non-linear dependence on $U$ for SOCAL and lubricated surfaces (dashed and solid lines, respectively). To validate the measurements using the cantilever force sensor, velocity data (open squares) of water droplets sliding down the same SOCAL surface at different $\tilt$ has been superimposed on the same plot. 

These observations account for the qualitatively different droplet motion on a tilted surface. A 10 $\mu$l water droplet was pinned on SH and SOCAL surfaces, when $\tilt$ is below a critical angle $\theta_{\text{crit}} \approx 5^{\circ}$; above $\theta_{\text{crit}}$, at $\tilt = 15^{\circ}$, the water droplet accelerated at $a$ = 0.4 m/s$^{2}$ on the SH surface, but moved at constant velocity $U_\text{const}$ = 8 mm/s on the SOCAL surface (Fig.~\ref{fig:droplet_motion}a, b). Eventually, the accelerating droplet on the SH surface will reach a terminal velocity---likely due air drag---but at a much larger $U_{\text{const}} \sim$ m/s \cite{reyssat2010dynamical}. In contrast, for lubricated surfaces, the droplet was never pinned and moved at increasing $U$ with increasing $\tilt$ (Fig.~\ref{fig:droplet_motion}c).

\begin{figure}
\includegraphics[]{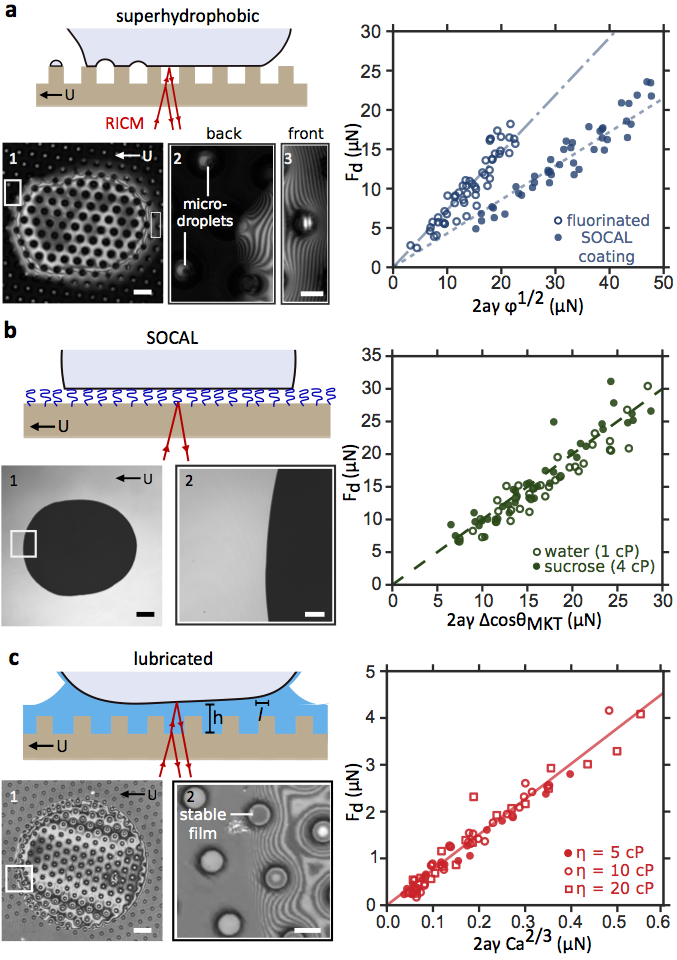} 
\caption{\label{fig:newton} Reflection interference contrast microscopy is used to visualize a) the intercalated air film on SH surface, b) the contact line on SOCAL surface, and c) the intercalated lubricant film on lubricated surfaces. Scale bars are 70 $\mu$m for a-1 and c-1, 20 $\mu$m for a-2,3 and c-2, 100 $\mu$m for b-1, and 30 $\mu$m for b-2. The dissipative forces $F_{d}$ are well-described by equations (\ref{eq:F_SH})--(\ref{eq:F_SLIPS}). $\Delta F_{d}$ is 1 $\mu$N for a,b and 0.2 $\mu$N for c.}
\end{figure}

 To understand the origin and hence the functional form of $F_{d}$, we analyzed the base of moving droplets on different surfaces using reflection interference contrast microscopy (RICM) (Fig.~\ref{fig:newton}) \cite{de2015air}. We used a similar set-up previously to study the lubricant dynamics of lubricated surfaces (See supplementary section S2) \cite{Dan2016oleoplaning}. Using RICM, we were able to confirm the presence of a stable micron-thick air film beneath the droplet on a SH surface (Cassie-Baxter state) and to visualize the details of the contact line with much improved temporal and spatial resolutions compared to other techniques. For example, previous studies using confocal \textit{fluorescence} microscopy usually require a dye to be added to the water droplet---which can affect its wetting properties---and can only achieve a temporal resolution of $\Delta t$ of several seconds \cite{schellenberger2016water}. Environmental Scanning Electron Microscopy (SEM) can achieve sub-micron spatial resolution, but again with poor $\Delta t$ of about 1s \cite{smyth2013visualization}. Moreover, the high-vacuum and low-temperature conditions of SEM may introduce artefacts and change the viscosity of the liquid(s), which in turn affects droplet behavior \cite{richard1999viscous, Dan2016oleoplaning}. 
 
Here, using RICM, we visualized the base of a droplet (without dye) moving on a transparent micropillar surfaces with a much improved $\Delta t <$ 0.1 s and good spatial details (Fig.~\ref{fig:newton}a-1). For example, the distortion of the contact line and the formation of capillary bridges on the micropillars at the receding front can be observed unambiguously (Fig.~\ref{fig:newton}a-1,2); we were also able to capture details such as micro-droplets that are left behind after the break-up of the capillary bridges, which then evaporate away (See Supplementary Figs S2--S3, and Supplementary Movie S1). In contrast to the distortion observed at the receding front, the advancing contact line was smooth and continuous (Fig.~\ref{fig:newton}a-3). Most of the pinning therefore occurs at the receding front, consistent with previous reports \cite{schellenberger2015direct, reyssat2009contact, gao2006contact}.    

We can estimate $F_{d}$ by assuming that the force due to each pillar $\sim \gamma d$ and the number of pillars in contact at the receding front $\sim 2a/d$:
\begin{equation} \label{eq:F_SH}
\begin{split}
F_{d} \sim (2a/p) \gamma d \approx 2a\gamma \phi^{1/2}
\text{,}
\end{split}
\end{equation}
where $\phi$ is the solid surface fraction. We confirmed this scaling law experimentally for water droplets of volumes $V$ = 0.5--8 $\mu$l moving at $U$ = 0.2--0.5 mm/s on different micropillar surfaces with different solid fractions $\phi$ = 0.05--0.23 ($d$ = 2--25 $\mu$m, $p$ = 5--50 $\mu$m, and $h_{p}$ = 5--30 $\mu$m). The prefactor in equation (\ref{eq:F_SH}) depends on the details of contact line distortion, which in turn depends on the surface functionalization; this explains the two different slopes observed in Fig.~\ref{fig:newton}a. The model described here, while simple, is able to account for the pinning force on SH surfaces, at least as well as other models previously proposed in the literature (see Supplementary Figs~S4 and S5) \cite{extrand2002model, joanny1984model, reyssat2009contact, butt2016energy}.  

Using RICM, we were also able to visualize the unique features of the contact line of a water droplet moving on a SOCAL surface at $U$ = 0.2 mm/s (Fig.~\ref{fig:newton}b). As was the case with SH surface, the shape of the contact line was elongated in the direction of the droplet motion, but unlike SH surface, the contact line at the receding front is smooth without any visibly discrete pinning points (Fig.~\ref{fig:newton}b-1,2, cf. Fig.~\ref{fig:newton}a-1,2). The functional form of $F_{d}$ for water and 30 wt$\%$ aqueous sucrose solution droplets of volumes $V$ = 1--5 $\mu$l moving at speeds $U$ = 0.1--1 mm/s is consistent with Molecular-Kinetic Theory (MKT):
\begin{equation} \label{eq:F_MKT}
\begin{split}
F_{d} &= 2a\gamma \dcos_{\text{MKT}} \\
	&= 2a\gamma [\dcos_{o} + 4 K_{B}T/\gamma \xi^{2} \arcsinh(U/2\Ko \xi)]
\text{.}
\end{split}
\end{equation}
In MKT, the movement of the contact line is modeled as an absorption-desorption process, with a series of small jumps of size $\xi$ and frequency $\Ko$, while $\dcos_{o}$ is $\dcos$ in the limit of $U \rightarrow 0$ (see Supplementary section S7)\cite{snoeijer2013moving, blake2006physics}. Viscous dissipation is unimportant, and $F_{d}$ is indistinguishable between water and 30 wt$\%$ sucrose droplets, despite their different viscosities, $\eta$ = 1 and 4 cP, respectively. Dashed line on Fig.~\ref{fig:newton}b shows the best-fit curve, with $\dcos_{o}$, $\xi$ and $\Ko$ as fitting parameters. The values obtained for $\xi$ = 3 nm and $\Ko$ = 7500 s$^{-1}$ are close to what were reported in the literature for other flat surfaces (see Supplementary Table S3) \cite{ranabothu2005dynamic, hayes1994molecular}. The value for $\dcos_{o}=0.07$, on the other hand, is much lower than typically encountered. For example, a flat glass or silicon surface rendered hydrophobic by fluorosilanization typically has $\theta_{\text{app}} = 110^{\circ}$ and $\hys$ = 15--30$^{\circ}$, or equivalently $\dcos_{o}$ = 0.3--0.5 \cite{quere1998drops}. The origin of the low $\dcos_{o}$ on SOCAL surfaces was hypothesized to originate from `liquid-like' polymer brushes that freely rotate at the moving contact line.

Interestingly, a combination of a SH and SOCAL surfaces, i.e. a micropillar surface coated with the same SOCAL polymer brush (filled circles, Fig.~\ref{fig:newton}a), behaves in a qualitatively different way from its flat SOCAL counterpart: $F_{d}$ no longer depends on $U$, and scales with equation (\ref{eq:F_SH}) rather than equation (\ref{eq:F_MKT}). Once again, this confirms that the pinning-depinning process at the micro-structured surface is fundamentally different from its chemically analogous flat surface. 

For lubricated surfaces with a stable, intercalated lubricant film, there is no contact line pinning and hence the droplet base is circular in shape and not elongated (Fig.~\ref{fig:newton}c-1,2, cf. Fig.~\ref{fig:newton}a-1,2 and Fig.~\ref{fig:newton}b-1,2). The entrainment of lubricant generates a hydrodynamic lift force, and the droplet levitates over the surface with a film-thickness given by the Landau-Levich-Derjaguin law, i.e. $h \sim R Ca^{2/3}$, where $Ca = \eta_{o}U/\gamma_{lo}$ is the capillary number, $\eta_{o}$ is the viscosity of the lubricant oil, and $\gamma_{lo}$ is the liquid droplet-lubricant-oil interfacial tension \cite{Dan2016oleoplaning, Landau1942}. $F_{d}$ is dominated by the viscous dissipation at the rim of the droplet's base of size $l \sim R Ca^{1/3}$, and is therefore given by:   
\begin{equation} \label{eq:F_SLIPS}
\begin{split}
F_{d} \sim (\eta U/h) 2al
	  \approx 2a \gamma_{lo} Ca^{2/3}
\text{.}
\end{split}
\end{equation} 
This was recently experimentally verified (reproduced here in Fig.~\ref{fig:newton}c for completeness) for droplets of $V$ = 1--5 $\mu$l moving at $U = 0.01$--5 mm/s, and with silicone oil of $\eta$ = 5--20 cP as lubricants \cite{Dan2016oleoplaning}. Note that this discussion is true only in the absence of solid-droplet contact; if for some reason, the lubricant film becomes unstable, $F_{d}$ becomes dominated by contact line pinning and is independent of $U$, reminiscent of the contact line pinning observed in SH surfaces (see Supplementary Fig.~S6).  

\begin{table}[h!]
\caption{\label{tab:hys} Nature of contact angle hysteresis}
\begin{ruledtabular}
\begin{tabular}{ccc}
	Surface &$\dcos$ &Comments\\
	\hline
	superhydrophobic &$\sim \phi^{1/2}$ &no dependence on $U$\\
	SOCAL &$\dcos_{o} + 4 K_{B}T/\gamma \xi^{2}$ &$\dcos \rightarrow \dcos_{o}$,\\ 
	      &$\arcsinh(U/2K\xi)$ & $U \rightarrow 0$\\ 
	lubricated &$ \sim Ca^{2/3}$ &$\dcos \rightarrow 0$, $U \rightarrow 0$
\end{tabular}
\end{ruledtabular}
\end{table} 

Comparing equations~(\ref{eq:F_SH})--(\ref{eq:F_SLIPS}) with equation~(\ref{eq:hys}), we can now get an expression for $\dcos$ for the different liquid-repellent surfaces, as summarized in Table~\ref{tab:hys}: $\dcos \sim \phi^{1/2}$ for SH surfaces, $\dcos = \dcos_{o} + 4 K_{B}T/\gamma \xi^{2} \arcsinh(U/2K\xi)$ for SOCAL surfaces, and $\dcos \sim (\gamma_{lo}/\gamma) Ca^{2/3} \approx Ca^{2/3}$ for lubricated surfaces. Note that for SOCAL and SH surfaces which have TPCL, $\dcos > 0$ when $U \rightarrow 0$; in contrast, for lubricated surfaces with no TPCL, $\dcos \rightarrow 0$ when $U \rightarrow 0$. Recently, there has been some debate on the correct physical interpretation of contact angle hysteresis for lubricated surfaces \cite{schellenberger2015direct, semprebon2017apparent, Dan2016oleoplaning}. We will address this more fully in a future publication, but in general $\dcos \sim Ca^{2/3}$ corresponds to optical measurements of macroscopic $\cos \trec - \cos \tadv$, and Furmidge's relation in equation (\ref{eq:hys}) can be applied to lubricated surfaces with some modifications (see Supplementary Fig.~S7).   

In summary, we have clarified the physics behind the three classes of liquid-repellent surfaces, in particular highlighting their distinct and unique properties, which are not captured by conventional contact angle measurements. We measured the dissipation force $F_{d}$ with sub-$\mu$N accuracy, and explicitly showed how the different functional forms of $F_{d}$ (and hence the corresponding contact angle hysteresis) arise from details of the contact line. While we have confined our discussion to liquid-repellency, many of the ideas and techniques outlined here are relevant to various other problems, ranging from ice-repellency to the rational design of non-fouling materials.   

\textbf{Acknowledgements}. The work was supported by the Office of Naval Research, U.S. Department of Defense, under MURI Award No. N00014-12-1-0875.  



\providecommand{\noopsort}[1]{}\providecommand{\singleletter}[1]{#1}%

\end{document}